\documentclass[aps,prl,twocolumn,showpacs,preprintnumbers,amssymb]{revtex4}

\usepackage{graphicx}

\let\bm=\bibitem

\def\ft#1#2{{\textstyle{{\scriptstyle #1}\over {\scriptstyle #2}}}}
\def\fft#1#2{{#1 \over #2}}

\def\del{\partial}
\def\nn{\nonumber}
\def\sst#1{{\scriptscriptstyle #1}}
\def\0{{\sst{(0)}}}
\def\1{{\sst{(1)}}}
\def\2{{\sst{(2)}}}
\def\3{{\sst{(3)}}}
\def\4{{\sst{(4)}}}
\def\5{{\sst{(5)}}}
\def\6{{\sst{(6)}}}
\def\7{{\sst{(7)}}}
\def\8{{\sst{(8)}}}
\def\ep{{{\epsilon}}}
\def\+{{{\sst +}}}
\def\-{{{\sst -}}}
\def\wedget{{{\scriptstyle\wedge}\,}}
\def\one{{{\sst 1}}}
\def\two{{{\sst 2}}}
\def\im{{\rm i}}

\newcommand{\bea}{\begin{eqnarray}}
\newcommand{\eea}{\end{eqnarray}}
\newcommand{\be}{\begin{equation}}
\newcommand{\ee}{\end{equation}}

\def\Bbb{\mathbb}
\def\R{{\Bbb R}}

\def\sim{{$\frak{sim}$}}
\def\isim{{$\frak{isim}$}}
\def\disim{{$\frak{disim}$}}

\begin{document}

\preprint{DAMTP-2007-68\ \ \ \ UB-ECM-PF-07-17\ \ \ \
MIFP-07-18\ \ \ \
}

\title{General Very Special Relativity is Finsler Geometry}

\author{G.W. Gibbons$^1$, Joaquim Gomis$^2$ and C.N. Pope$^{3}$}
\altaffiliation[]{Research supported in part
by DOE grant DE-FG03-95ER40917.}
\affiliation{%
${}^1\!\!\!$ DAMTP, Centre for Mathematical Sciences, Cambridge University
Wilberforce Road, Cambridge CB3 OWA, UK\\
${}^2\!\!\!$ Departament ECM, Facultat de F\' isica, Universitat de Barcelona,
Diagonal 647, E-08028 Barcelona, Spain\\
${}^3\!\!\!$ George P. \& Cynthia W.
Mitchell Institute for Fundamental Physics,
Texas A\&M University, College Station, TX 77843-4242, USA
}%

\date{July 13, 2007}

\begin{abstract}

   We ask whether Cohen and Glashow's Very Special
Relativity model for Lorentz violation might be modified, perhaps
by quantum corrections, possibly producing a curved spacetime with
a cosmological constant.  We show that its symmetry group
ISIM(2) does admit a 2-parameter family of continuous deformations, but
none of these give rise to non-commutative translations analogous
to those of the de Sitter deformation of the Poincar\'e group:
spacetime remains flat.   Only a
1-parameter family DISIM$_b$(2) of deformations of SIM(2)
is physically acceptable.  Since this could arise through quantum corrections,
its implications for tests of Lorentz violations via the Cohen-Glashow
proposal should be taken into account.  The Lorentz-violating point particle
action invariant under DISIM$_b$(2) is of Finsler type, for which the line
element is homogeneous of degree 1 in displacements, but anisotropic.
We derive DISIM$_b$(2)-invariant
wave equations for particles of spins 0, $\ft12$ and 1.
The experimental bound, $|b|<10^{-26}$, raises the question ``Why is the
dimensionless constant $b$ so small in Very Special Relativity?''

\end{abstract}

\pacs{11.25.-w, 98.80.Jk, 04.50.+h}
\maketitle

   Local Lorentz and CPT invariance are fundamental assumptions in almost all
current physical theories.  It is important to test these assumptions
experimentally, lest evidence of new physics beyond the standard model
be overlooked.  Current experimental limits on violations of local
Lorentz and CPT invariance are extremely stringent.  Thus what is required are
novel alternative non-Lorentz invariant theories, capable of circumventing
these tight limits.
Recently, Cohen and Glashow \cite{cohgla} have made the ingenious
proposal that the
local laws of physics need not be invariant under the full Lorentz
group, generated by $M_{\mu\nu}$, but rather, under a SIM(2) subgroup,
whose Lie algebra is
generated by $(M_{\+i}, M_{ij}, M_{\+\-}=M_{{\scriptstyle 03}},
)$ (with $i$ and $j$ ranging over
the values 1 and 2) \cite{foot1}.
    This they referred to as {\it Very Special Relativity}.
Taking the semi-direct product with the translations $(P_\+,P_\-,P_i)$ gives
an 8-dimensional subgroup of the Poincar\'e group called ISIM(2)
\cite{kogper}.

  The
great merits of Cohen and Glashow's suggestion are that CPT symmetry is
preserved and that ISIM(2) leaves
invariant no vector or tensor fields, known as ``spurion
fields.'' For example, a spurionic vector field may be thought of as the
4-velocity of the {\ae}ther \cite{Dirac}.  In fact SIM(2) consists of those
Lorentz transformations $\Lambda^\mu{}_\nu$ leaving invariant the null
direction $n^\mu=\delta^\mu_+$, i.e. such that $\Lambda^\mu{}_\nu n^\nu=
\lambda n^\mu$ for some $\lambda$ which depends on $\Lambda^\mu{}_\nu$.
The generator $M_{\+\-}$ acts by sending
$n^\mu\rightarrow
\lambda n^\mu$.  This scaling symmetry implies that
one cannot take $n^\mu$ to define the actual 4-velocity of the
{\ae}thereal motion, but only its direction, thus rendering the
presence of such an
{\ae}ther more difficult to detect.  A theory of this kind appears to be
compatible with all current
experimental limits on violations of Lorentz invariance and spatial
isotropy \cite{cohgla,dume}.

  Subsequently, Cohen and Freedman \cite{cohfre}, and later
Lindstr\"om and Ro\v cek \cite{linroc}, showed that ISIM is
compatible with supersymmetry. There are several ways one might try
to incorporate gravity.  One is where we make local the  ISIM(2)
algebra \cite{cgq}. Another is to consider a global space-time and
to see if it is compatible with very special relativity ideas. In
this case there appear to be difficulties \cite{frepriv}.   For
example, the maximal subgroup of $SO(4,1)$, the isometry group of de
Sitter spacetime, is SIM(3), which is 7-dimensional \cite{patera}.
SIM(3) contains SIM(2) as a subgroup, but the stabilizer, or tangent
space group, is SO(3), not SIM(2).

  Alternatively, we recall that
that Poincar\'e group admits a unique
deformation into the de Sitter (anti-de Sitter) group \cite{levynahas},
with $[P_\mu,P_\nu]= \ft13 \Lambda M_{\mu\nu}$, where
the parameter $\Lambda$ is the cosmological constant. One may
ask whether ISIM(2) admits a similar
deformation, such that the translations $P_\mu$ become non-commutative.  If
so, the coset of the deformed group divided by SIM(2) (or its deformation)
could be thought of as a curved spacetime.

   Here we show that there are indeed continuous
deformations of ISIM(2), but in all of them the translations remain
commutative.  Among them is a 1-parameter family of deformations,
which we denote by DISIM$_b$(2). For any values of $b$ this is an
8-dimensional subgroup of the 11-dimensional Weyl group, i.e. the
semi-direct product of dilatations with the Poincar\'e group.
Subgroups with different values of $b$ are not isomorphic.
Interestingly, if one constructs a point-particle action for the
deformed groups DISIM$_b$(2), using the methods of non-linear
realisations \cite{colweszum}, one arrives at Lagrangians of Finsler
form, first proposed by Bogoslovsky (see \cite{bog} and references
therein).  Therefore the
deformation of very special relativity leads in a natural way
to Finsler geometry. In the remainder of this letter we shall
outline the derivation of these results, and comment on their
physical significance.

   Continuous deformations of Lie algebras have been extensively explored,
by both mathematicians and physicists, under the rubric of Lie-algebra
cohomology \cite{levynahas}.  Here we give an elementary account
based on the Cartan-Maurer equations, which provides a simple
and easily automated scheme for determining the deformations of
a given Lie algebra $\mathfrak g$ with structure constants $C_a{}^b{}_c$.
We suppose there exists a family of deformed Lie algebras ${\mathfrak g}_t$
parameterised by a continuous variable $t$, with structure constants
\be
\hat C_a{}^b{}_c(t) = C_a{}^b{}_c + t\, A_a{}^b{}_c + t^2\, B_a{}^b{}_c
            +\cdots\,.\label{Cdef}
\ee
We are only interested in deformations which do not arise merely from a
($t$-dependent) change of basis:
\be
\hat C_a{}^b{}_c(t) = S^b{}_e C_d{}^e{}_f (S^{-1})^d{}_a\, (S^{-1})^f{}_c
\,,\quad S^a{}_b\in GL(n,\R)\,.\nn
\ee
Expanding the Jacobi identity
$\hat C_d{}^e{}_{[a}(t)\, \hat C_b{}^d{}_{c]}(t)=0$ in powers of $t$ gives
rise at linear order to
\be
C_d{}^e{}_{[a}\,  A_b{}^d{}_{c]}+ A_d{}^e{}_{[a}\,  C_b{}^d{}_{c]}=0\,.
\label{fodef}
\ee
A first-order deformation $A$ will be trivial if $S^a{}_b(t)= \delta^a_b +
  t\, \Phi^a{}_b+\cdots $ and
\be
A_a{}^b{}_c = \Phi^b{}_e\, C_a{}^e{}_b - C_e{}^b{}_c\, \Phi^e{}_a -
   C_a{}^b{}_e\, \Phi^e{}_c\,.
\ee

   Introducing a basis $\lambda^a$ of left-invariant 1-forms of the
original algebra,
such that $d\lambda^a= -\ft12 C_b{}^a{}_c\, \lambda^b\wedget \lambda^c$, we
define vector-valued 1-forms and 2-forms $\Phi^a\equiv \Phi^{}_b\,
\lambda^b$ and $A^a\equiv \ft12 A_b{}^a{}_c\, \lambda^b\wedget \lambda^c$
and a matrix-valued 1-form $C^a{}_b \equiv \lambda^c C_c{}^a{}_b$.
Defining $D\equiv d+C\wedget$, the first-order deformation equations
(\ref{fodef}) may then be written as
\be
DA =0\,,\qquad A \ne -D\Phi\,,
\label{nontriv}
\ee
where the second equation expresses the requirement of non-triviality of
the deformation.  Because $D^2=0$ as a consequence of the Jacobi identities
of the undeformed algebra, $dC+ C\wedget C=0$, the differential $D$ may be
regarded as a co-boundary operator acting on $\mathfrak g$-valued forms.
The non-trivial linearised deformations are therefore in 1-1
correspondence with the second cohomology group
$H^2(\mathfrak g; \mathfrak g)$.

   If a non-trivial linear deformation $A$ is found, the next step is to
investigate the Jacobi identities at order $t^2$. These read
\be
C_d{}^e{}_{[a}\,  B_b{}^d{}_{c]}+ B_d{}^e{}_{[a}\,  C_b{}^d{}_{c]}
  + A_d{}^e{}_{[a}\,  A_b{}^d{}_{c]}=0\,,
\label{sodef}
\ee
and can be re-expressed in terms of the vector and matrix valued forms as
\be
DB + A\bullet A=0\,,\quad (A\bullet A)^e \equiv
  \ft12 A_d{}^e{}_{[a}\,  A_b{}^d{}_{c]}
  \lambda^a\wedget \lambda^b\wedget\lambda^c\,.\label{DB}
\ee
This equation can only be solved if $D(A\bullet A)=0$, which implies that
$A\bullet A$ must be both $D$-closed and exact.  Thus there is a potential
obstruction to finding a deformation at quadratic order: $A\bullet A$ should
not have a projection in the third cohomology group
$H^3(\mathfrak g;\mathfrak g)$.  There
are analogous equations to (\ref{DB}) at higher orders in $t$.  If
$H^3(\mathfrak g;\mathfrak g)$ vanishes, then the equations may be solved
at all non-linear orders.  If $H^3(\mathfrak g;\mathfrak g)$ is non-zero,
then the higher-order analogues of $ (A\bullet A)$ should have no
projections into it.

   We start with the ISIM(2) Cartan-Maurer relations
\bea
d\lambda^\+ &=&\lambda^{\+i}\wedget \lambda^i + \lambda^{\+\-}\wedget
\lambda^\+\,,\qquad
d\lambda^\- = -\lambda^{\+\-}\wedget \lambda^\- \,,\nn\\
d\lambda^i&=& \ep_{ij} \lambda^{\one\two}\wedget \lambda^j
     + \lambda^\-\wedget \lambda^{\+ i} \,,\nn\\
d\lambda^{\+ i}&=& \ep_{ij} \lambda^{\one\two}\wedget
     \lambda^{\+ j} +\lambda^{\+\-}\wedget
                 \lambda^{\+ i}\,,\nn\\
d\lambda^{\+\-} &=&0\,,\quad d\lambda^{\one\two}=0\,,\label{formalg}
\eea
where $g^{-1}dg= \lambda^\+ P_\+ \,+\, \lambda^\- P_\- \,+\, \lambda^i P_i
\,+\, \lambda^{\+ i} M_{\+i} \,+\, \lambda^{\+\-} M_{\+\-} \,+\,
 \lambda^{\one\two} M_{\one\two}$.
Defining $N\equiv M_{\+\-}$ and $J\equiv M_{\one\two}$,
the corresponding non-trivial Lie brackets are therefore
\bea
&&[N,P_\pm]= \mp P_\pm\,,\quad
[N,M_{\+ i}] = -M_{\+ i}\,,\nn\\
&& [J,P_i]=\ep_{ij} P_j
\,,\quad [J,M_{\+ i}]=\ep_{ij} M_{\+ i}\,,\nn\\
&&[M_{\+ i},P_\-]=P_j\,,\quad
[M_{\+ i},P_j] = -\delta_{ij} P_\+\,.
\eea

  Expanding the vector-valued 2-form $A^a$ on a basis of 2-forms and solving
the resultant linear equations in (\ref{nontriv}) reveals that there is
a 2-parameter family of non-trivial solutions, i.e.,
$H^2(\hbox{\isim(2)};\hbox{\isim(2)})$
is 2-dimensional.   Substituting this
linearised solution into the full Jacobi identities, we find that it gives
a 2-parameter family of {\sl exact} Lie algebras of the leading-order
form (\ref{formalg}), with additional terms as follows:
\be
d\lambda^\mu \longrightarrow d\lambda^\mu +
  a \lambda^{\one\two}\wedget \lambda^\mu
         + b \lambda^{\+\-}\wedget \lambda^\mu\,,
\ee
where $\mu=(\+,\-,i)$. Here $a$ and $b$ are arbitrary constant parameters.

  As in the undeformed case, the algebra here has the structure of a
semi-direct sum of \sim(2) and the translations $\R^4$, i.e. \sim(2)$\ltimes
  \R^4$.  While the
$M_{\+ i}$ act on the translations as in the undeformed case, the adjoint
action of the generators $N$ and $J$ is given by $[N,P_\mu]= P_\nu\,
C_N{}^\nu{}_\mu$ and $[J,P_\mu]= P_\nu\, C_J{}^\nu{}_\mu$, where the matrices
$C_N$ and $C_J$ are given respectively by
\bea
-\pmatrix{b+1 & 0& 0& 0\cr
                    0 & b-1 &0 &0\cr
                    0 & 0 & b & 0\cr
                    0& 0 & 0 & b }\,,\
\pmatrix{ -a & 0& 0& 0\cr
                    0 & -a &0 &0\cr
                    0 & 0 & -a & -1\cr
                    0& 0 & 1 & -a }\,.\nn
\eea
Minkowski spacetime may be thought of as the symmetric space
$E(3,1)/SO(3,1)$ with $SO(3,1)$ playing the role of the
tangent-space group, and the tangent space being spanned by the
translations $P_\mu$. In our case we wish to replace $SO(3,1)$ by
SIM(2). However, it follows by exponentiating $C_J$ that
$J$ does not generate a compact $SO(2)$ subgroup unless the deformation
parameter $a$ vanishes.
From now on we shall restrict attention to this $a=0$
case, for which we denote the deformed algebra by \disim$_b$(2).

  The non-trivial Lie brackets for \disim$_b$(2) are given
by
\bea
&&[N,P_\pm]= -(b\pm1) P_\pm\,,\quad [N,P_i]=-b P_i\,,\nn\\
&&[N,M_{\+ i}] = -M_{\+ i}\,,\quad [J,P_i]=\ep_{ij} P_j
\,,\nn\\
&&[J,M_{\+ i}]=\ep_{ij} M_{\+ i}\,,\quad [M_{\+ i},P_\-]=P_j\,,\nn\\
&&[M_{\+ i},P_j] = -\delta_{ij} P_\+\,,
\eea
The deformed group DISIM$_b$(2) is a subgroup of the Weyl group with
an action on Minkowski spacetime given by translations, and boosts
in the ${\scriptstyle +} i$ directions, together with a combination
of a boost in the ${\scriptstyle +-}$ direction and a dilatation.
Specifically, the deformed generator N acts as
\be
x^i\rightarrow \lambda^{-b}\, x^i\,,\quad
x^\-\rightarrow \lambda^{1-b}\, x^\-\,,\quad
x^\+\rightarrow \lambda^{-1-b}\, x^\+\,.\label{scalings}
\ee

The group DISIM$_b$(2) does not leave invariant the standard
Minkowski line element $ds=(\eta_{\mu\nu} dx^\mu dx^\nu)^{1/2}$, but
rather, the Finslerian line element
\bea
ds &=& (2dx^\+ dx^\- + dx^i dx^i)^{(1-b)/2}\, (dx^-)^b\,,\nn\\
&=& (\eta_{\mu\nu} dx^\mu dx^\nu)^{(1-b)/2}\, (n_\rho dx^\rho)^b\,.
\eea
This is of the form first suggested by Bogoslovsky \cite{bog}.

  We shall now construct, using the theory of non-linear realisations, a
DISIM$_b$(2)-invariant Lagrangian for a point particle.  We
parameterise the coset DISIM$_b$(2)$/SO(2)$ as
\be
g=e^{x^\mu P_\mu} e^{w^i M_{\+i}} e^{w N}\,,\label{cosetrep}
\ee
which implies that
\bea
g^{-1} dg &=& dw N + e^w dw^i M_{\+i} + e^{-w(1-b)} dx^\- P_\- \nn\\
&&+e^{w(1+b)} (dx^\+ + w^i dx^i - \ft12 w^i w^i dx^\-) P_\+\nn\\
&&+ e^{wb} (dx^i-w^i dx^\-) P_i\,,\\
&=& \lambda^{\+\-} N + \lambda^{\+i} M_{\+i} + \lambda^\- P_\- +
   \lambda^\+ P_\+ + \lambda^i P_i\,,\nn
\eea
where $(\lambda^{\+\-},\lambda^{\+i},\lambda^{\+}, \lambda^\-,\lambda^i)$ are
the restrictions (or pullbacks) of the invariant 1-forms on the group
DISIM$_b$(2) to the coset DISIM$_b$(2)$/SO(2)$.

  In order to construct a DISIM$_b$(2)-invariant Lagrangian with worldline
reparameterisation invariance, we allow the Goldstone coordinates
$(w,w^i,x^\mu)$ to depend on the worldline coordinate $\tau$ (see, for example,
\cite{gomkamwes}).  We
shall restrict our attention to Lagrangians that are
linear in the left-invariant 1-forms pulled back to the particle's
worldline.  Requiring invariance under $SO(2)$
then implies that we must discard $\lambda^{+i}$ and $\lambda^i$,
and thus we consider the Lagrangian
\be
{\cal L}= \alpha e^{w(1+b)} (\dot x^\+ + w^i \dot x^i -\ft12 w^i w^i
        \dot x^\-) + \beta e^{-w(1-b)} \dot x^\- + \gamma \dot w\,,
\label{lag}
\ee
where $\alpha$, $\beta$ and $\gamma$ are arbitrary constants. Since the
last term is a total derivative, we can discard it.  Eliminating the
non-dynamical Goldstone coordinates $w$ and $w^i$, one obtains, in
physical units, the Lagrangian
\be
{\cal L}= -m  (-\eta_{\mu\nu} \dot x^\mu \dot x^\nu)^{(1-b)/2}\,
     (-n_\rho \dot x^\rho)^b\,.\label{lag1}
\ee

   Calculating th canonical momenta from (\ref{lag1}), we
obtain the DISIM$_b$(2)-invariant dispersion relation or Hamiltonian
constraint
\be
\eta^{\mu\nu} p_\mu p_\nu=- m^2(1-b^2)\Big(-\fft{n^\nu p_\nu}{m(1-b)}
       \Big)^{2b/(1+b)}\,.
\ee
Note that that for $b=0$ we recover the ordinary free
relativistic particle, which does not see the
light-like direction $n^\mu$.  The cases $b=\pm1$ are special, and
are best investigated directly
from equation (\ref{lag}).  The conclusion is that for $b=1$ we obtain the
massless equation $\eta^{\mu\nu} p_\mu p_\nu=0$, whilst for $b=-1$ we have
$\dot x^i=0$ and $\dot x^-=0$, and the dynamics is trivial in this case.

Upon quantisation, $p_\mu\rightarrow -\im \del_\mu$, we obtain a generalised
Klein-Gordon equation of the form
\be
-\square \phi + m^2 (1-b^2)
\Big(\fft{\im n^\mu\del_\mu}{m(1-b)}\Big)^{2b/(1+b)} \phi=0\,.
\label{genkg}
\ee
This is in general a non-local equation, since it involves fractional
derivatives.  Although the special case $b=1$ appears to give a local
modification of the usual Klein-Gordon equation involving a term
linear in $n^\mu\del_\mu$ this is really equivalent to the standard
massless Klein-Gordon equation (as discussed earlier for this special
value of $b$).  Specifically, the first-order term can be removed by making
the phase transformation $\phi\rightarrow \phi\, e^{-\im m n_\mu x^\mu/2}$.

   The free Maxwell equations are also invariant under the action of
the Weyl group and so they too are clearly invariant under DISIM$_b$(2).
The invariance of $A_\mu dx^\mu$, together with (\ref{scalings}), implies that
$(A_\+,A_\-,A_i)\rightarrow (\lambda^{b+1}A_\+,
\lambda^{b-1} A_\-,\lambda^b A_i)$.  Since $d^4x \rightarrow \lambda^{-4b}
d^4x$ any invariant action must have ${\cal L}\rightarrow \lambda^{4b}
{\cal L}$.  Thus we can add a mass term, giving
\be
{\cal L}=-\ft14 F_{\mu\nu} F^{\mu\nu}
  -\ft12 m^2 \Big(\fft{(n^\mu A_\mu)^2}{A_\nu A_\nu}\Big)^b
          A_\rho A^\rho\,.
\ee
If  $b=1$ we can further include a non-Lorentz invariant
Chern-Simons term \cite{harleh}$, {\cal L}_{\hbox{cs}}= \ft12\ell^{-1}\,
     \epsilon^{\mu\nu\rho\sigma} n_\mu A_\nu F_{\rho\sigma}$,
where $\ell$ is an arbitrary length scale.

   Since it is a subgroup of the Weyl group, DISIM$_b$(2) leaves invariant
the massless Dirac Lagrangian.
    Bogoslovsky and Goenner \cite{boggoe} have pointed out that adding
to the massless Dirac Lagrangian a term of the form
\be
m\, \Big[\Big( \fft{\im n_\mu \bar\psi \gamma^\mu\psi}{\bar\psi \psi}\Big)^2
  \Big]^{b/2}\, \bar\psi\psi
\ee
gives a non-linear DISIM$_b$(2)-invariant generalisation of the massive
Dirac equation.  This follows from the scalings
$\psi\rightarrow \lambda^{3b/2}\psi$,\
$\gamma^\mu\del_\mu\rightarrow \lambda^b \gamma^\mu\del_\mu$ under the
action of the generator $M_{\+\-}$.  As with the generalised Klein-Gordon
equation (\ref{genkg}), the case $b=1$ is special: The additional term
may then be removed by a phase transformation of the form
$\psi\rightarrow \psi\, e^{\im m n_\mu x^\mu}$.  Note however that, as
discussed below, experimental bounds constrain $|b|$ to be very much
less than 1.

   CPT will be preserved if an operator exists in the complexification of
DISIM$_b$(2) that reverses $x^\mu$.
As discussed for ISIM(2) in \cite{cohgla}, a candidate CPT operator
is $e^{\im\phi J} e^{\im \alpha N}$.  This has the following action on
the momenta:
\be
(P_\+,P_\-,P_\one + \im P_\two)\rightarrow e^{-b\alpha}\,
(e^{-\alpha} P_\+, e^{\alpha}P_\-, e^{\im\phi}(P_\one + \im P_\two))\,.
\ee
Requiring that $P_\mu\rightarrow -P_\mu$ implies that
\be
\alpha=\im\pi(n_\- - n_\+) \,, \phi=\pi(2n_{\sst 3} -n_\+ - n_\-)
\,, b= \fft{1+n_\+ + n_\-}{n_\+ - n_\-}\,,
\ee
for integers $n_\+$, $n_\-$ and $n_{\sst 3}$.  Although $b=1+p/q$ is rational,
with $p$ odd, one may always choose $n_\+$
and $n_\-$ so that $b$ is arbitrarily close to any given real number.

   We have shown that ISIM(2) admits no deformations
with de Sitter-like non-commutative translations. However, it is
interesting to note that ISIM(2), unlike the Poincar\'e group but like
the Galilei group, admits
a central extension: the
cohomology group $H^2(\hbox{\isim(2)},\R)$ is non-trivial. We find that it is
generated by $\lambda^{\+\-}\wedget \lambda^{\one\two}$, and so may adjoin
to \isim(2) a central element $Z$, whose only non-trivial Lie bracket is
\be
[N,J]= Z\,.
\ee
Thus unlike the translations, the boosts and rotations can be
rendered non-commutative.
This also works for the full 2-parameter family of deformations
of ISIM(2).  Including the extra generator $Z$, appending $e^{\theta Z}$,
and proceeding with the comstruction of an invariant particle Lagrangian
leads to unmodified equations of motion, since the only effect is to add
a total derivative $\dot\theta$ to the Lagrangian.

    It was argued in \cite{bog} that {\ae}ther-drift experiments imply
$|b|<10^{-10}$.  However, it follows from (\ref{lag1}) that every
particle has a mass tensor
$m_{ij}=  (1-b) m (\delta_{ij} + b n_i n_j)$.  Hughes-Drever type limits
\cite{hudr} on the
anisotropy of inertia then potentially
imply that $|b|<10^{-26}$.  However, this depends on the precise form
of the interactions \cite{bogo2,fangolski}.  Since a non-vanishing
$b$ could arise through quantum corrections, Very Special Relativity faces
the question, analogous to the puzzle posed by the cosmological
constant in traditional relativity: ``Why is $b$ so small?''

\bigskip


\begin{acknowledgments}

  We are grateful to  
Dan Freedman and Jaume Gomis for helpful discussions,
and  to the Galileo Galilei Institute in Florence, the workshop on
{\it String and M-Theory Approaches to Particle Physics and
Cosmology}, and the INFN, for support and hospitality during the
course of this work.  We thank George Bogoslovsky for a helpful discussion
about upper bounds on the parameter $b$.  This work has also been
supported by  the European EC-RTN project MRTN-CT-2004-005104, MCYT
FPA 2004-04582-C02-01,  CIRIT GC 2005SGR-00564.


\end{acknowledgments}

\end{document}